# Hybrid control strategy for a semi active suspension system using fuzzy logic and bio-inspired chaotic fruit fly algorithm

**Vikram Bhattacharjee**
Department of Electrical & Computer Engineering
Carnegie Mellon University
Email:vbhattac@andrew.cmu.edu

**Debanjan Chatterjee**
Department of Electrical Engineering
Jadavpur University
Email:debanjanee84@gmail.com

**Orkun Karabasoglu***
Sun Yat-sen University-Carnegie Mellon University (SYSU-CMU)
Joint Institute of Engineering, Guangzhou, China
SYSU-CMU Joint Research Institute, Shunde, China
Carnegie Mellon Scott Institute for Energy Innovation
Carnegie Mellon University, Pittsburgh, PA, US
Email: karabasoglu@cmu.edu

## ABSTRACT

*This study proposes a control strategy for the efficient semi active suspension systems utilizing a novel hybrid PID-fuzzy logic control scheme .In the control architecture, we employ the Chaotic Fruit Fly Algorithm for PID tuning since it can avoid local minima by chaotic search. A novel linguistic rule based fuzzy logic controller is developed to aid the PID.A quarter car model with a non-linear spring system is used to test the performance of the proposed control approach. A road terrain is chosen where the comfort and handling parameters are tested specifically in the regions of abrupt changes. The results suggest that the suspension systems controlled by the hybrid strategy has the potential to offer more comfort and handling by reducing the peak acceleration and suspension distortion by 83.3 % and 28.57% respectively when compared to the active suspension systems. Also, compared to the performance of similar suspension control strategies optimized by stochastic algorithms such as Genetic Algorithms (GA), Particle Swarm Optimization (PSO) and Bacterial Foraging Optimization (BFO), reductions in peak acceleration and suspension distortion are found to be 25%, 32.3%, 54.6% and 23.35 %, 22.5%, 5.4 % respectively.The details of the solution methodology have been presented in the paper.*

*Keywords:* Semi-Active Suspension; Chaotic Fruit Fly Algorithm; PID Tuning, Fuzzy Logic

# 1. Introduction

Suspension systems provide vehicular stability and ride comfort during transportation by reducing vibrations arising due to random variations in road elevation from the reference level. Development of adaptive approach to counter such adverse changes in the external conditions, has been a topic of interest among researchers down the years. The suspension systems in vehicles can be categorized broadly into three types: (1) Passive, (2) Active and (3) Semi-Active Suspension System. The Passive Suspension System consists of a spring assembly to maintain system stability and has the lowest efficiency when compared to other systems. The Active System consists of an actuator to control the vibration in the system. However the Active Systems suffer from drawbacks like high power consumption and weight to power ratio. The Semi-Active Suspension Systems, on the other hand, utilize magneto-rheological dampers for control and require low implementation cost. The control architecture for a semi active suspension systems is simpler when compared to the active suspension and requires no modifications before installation. The MR Dampers work with the help of magneto-rheological fluids whose damping characteristics are a function of an applied magnetic field. Semi-active suspension systems although suffer from some disadvantages like operational constraints but lower performance with respect to active suspension systems. Appropriate estimation of the parameters characterizing the damper force and development of control architecture will help achieve high levels of ride comfort and road handling. Carlson et al. (1995) [1] presented the experimental model of a magneto-rheological damper (MR) for applications in suspension systems and demonstrated that controlling action of the damper forces could help increase ride comfort.

There has been some control applications for suspension systems using Genetic Algorithms [2,3], LQ Controllers [4,5], Fuzzy Logic Controllers [6,7] alongside with the PID tuning [8,9] to increase vehicular stability and ride comfort. The PID controllers depend only on the measured variables of the dynamical system and not on the physical knowledge of system and hence they are adopted widely in industry [10]. On the other hand, the fuzzy controllers provide a flexible method to account for the impact of interaction of the various system inputs by applying rule sets which are based on the occurrences of different events in the system [11]. Given the advantages of PID and rule based controllers, there has been two approaches for the implementation of Fuzzy-PID hybrids in a suspension system: (1) Tuning the PID parameters with fuzzy controllers or (2) using the control strategies independently in the system for achieving ride comfort and handling. Jinzhi et al. [12] devised a control strategy for suspension systems using a GA based PID tuning coupled with a fuzzy controller where the PID minimized vertical body acceleration and the fuzzy controller minimized the pitch acceleration. Shanfa et al. (2006) [13] designed a switching control strategy between a PID and a Fuzzy System in a semi active suspension system. The Fuzzy-PID control strategy reduced vertical acceleration and provided robust performance when compared to the individual Fuzzy and PID control strategies.

In this study we develop a Fuzzy-PID hybrid control strategy for MR Dampers in the semi active suspension systems. The PID controller has been utilized to suppress the suspension distortion and the Fuzzy Logic Controller supplies the variable damper force. In this paper, the Chaotic Fruit Fly Algorithm has been utilized for the tuning of the PID controller and the Fuzzy controller was designed considering the variation of the damper forces with suspension parameters such as distortion, sprung mass velocity and acceleration. The Chaotic Fruit Fly Algorithm [14] builds upon the Fruit Fly Algorithm [15] where the search process is improved by the incorporation of logistic mapping in order to avoid local optima. The adaptive Chaotic Fruit Fly algorithm has a more accurate and faster searching capability than other swarm intelligence algorithms and its familiarity with responding to the dynamics of the non-linear systems makes it an ideal choice for PID tuning in a semi active suspension system.

This paper is the first study that employs chaotic fruit fly algorithm in the domain of mechanical engineering. We compared ride comfort provided by our proposed strategy against the Active System and Fuzzy-PID systems tuned by several methods such as Ziegler Nichols, Genetic Algorithms (GA), Particle Swarm Optimization (PSO) and Bacterial Foraging Optimization (BFO). We demonstrated the improvement of the ride comfort parameters such as suspension distortion, sprung mass acceleration and the tire loading of the system.

## 2. Literature Review

Some research in the literature focuses on the development of the control strategies for magneto-rheological dampers in suspension systems using PID controllers with different tuning algorithms. Liu Wei et al. (2010) developed a PID Controller using Neural Networks for the control of MR Dampers [16]. The strategy improved the ride comfort in a semi active suspension system when compared to the passive suspension system. Ren et. al. (2008) used Neural Networks to devise a self-tuning strategy for a PID controller to govern the motion of a two wheeled vehicle. The experimental results proved that the control strategy could reduce instability during motion [17]. Dirman Hanafi (2010) designed a PID controller for a semi active suspension system where the development of the controller was based on a suspension model designed using System Identification [8]. The response of the control strategy matched the road input signal which proved that the controller could successfully control the motion and was capable of increasing ride comfort.

In addition to the PID controllers researchers have studied the scope of fuzzy logic controllers in the design of suspension systems to increase ride comfort and handling. They incorporated several parameters such as suspension distortion, vehicle speed and acceleration, which influence the performance of the suspension systems. He et al. (2010) developed a fuzzy controller based on the effects of change in the yaw rates and the side slips on the break, and throttle of the vehicle. The control strategy helped improve vehicle handling and increased ride comfort when compared to the passive suspension system [18]. Nicolas et al. (1997) adopted two different approaches to develop fuzzy logic controllers (1) based on the actions of the driver and (2) based on vehicle. The proposed strategy helped achieve performance similar to the semi active suspension system at lower sensor costs [19]. In [20] a fuzzy controller was designed for a semi active suspension system taking into consideration multiple rule sets. The fuzzy controller increased ride comfort and road holding by 28% and 31% respectively when compared to that of a linear quadratic regulator controlled semi-active suspension system. In [21] a fuzzy logic control system was developed for a semi-active suspension system where the control force is a function of body velocity and suspension velocity. The method showed improvement over the passive and the Skyhook control system.

The above cited papers indicate that both fuzzy logic and PID controllers have the capability of increasing vehicular stability and can adapt to changes in the system parameters due to the presence of irregular road disturbances. Based on these advantages researchers have studied the viability of this hybrid approach in suspension systems. In [22] a fuzzy based PID Controller is used in semi active control of the MR Dampers of a vehicle suspension system. The study revealed that the Fuzzy based PID Controllers can achieve a peak reduction of the frequency of vibration up to 83% of the natural frequency of vibration under no control. However minimization of the body acceleration and discomfort factors increases actuator forces and power consumption to a huge extent. Thus to prevent material damage efficient optimization algorithms should be used to tune PID controllers. In [23] a new PID tuning algorithm by the fuzzy set theory has been developed which reduces overshoots and rise time to an extent larger than the active and passive suspension systems.

In this paper we explore the application of a novel hybrid control strategy that integrates (1) the PID control, to reduce suspension distortion, and (2) Fuzzy Logic Control to provide the variable damper force. The PID controller is tuned with the adaptive Chaotic Fruit Fly algorithm and the Fuzzy Logic Controller is based on a linguistic rule set which takes into account the variability of damper forces on travel parameters like suspension distortion, body acceleration and body velocity. The performance of the proposed strategy has been compared with the Active Suspension System and PID systems tuned by Genetic Algorithms (GA), Particle Swarm Optimization (PSO) and Bacterial Foraging Optimization (BFO).

## 3. Methodology

### 3.1 The Semi Active Suspension System And The Quarter Car Model

The semi active suspension systems reduce power consumption and increases vehicular stability during transportation compared to the active suspension system. The system utilizes a damper whose characteristics can be changed externally using controllers in contrast to the fixed conventional active [24] and passive [25] systems. The design of the semi active system can be represented with a quarter car model [26] where the masses can be classified into sprung ($m_s$) and unsprung mass ($m_u$). The sprung mass is damped by the suspension system and the unsprung mass consists of the mass of the components of the vehicle system such as the brakes and the steering wheel. The system consists of non-linear springs between the sprung and the unsprung masses and are defined by the stiffness constants as described in the Figure 1. below. A damper also exists between the tire and the sprung mass with a stiffness constant $c_0$. However the overall contribution of the tire damper is often neglected due its negligible effect on the performance [27]. In a semi active system, a variable damper exists between the sprung and the unsprung masses which is controlled by an external control unit. A road profile $z_g$ through tire contact is used to represent irregularities of the road surface. The guiding equations of motion (Equation (1) and (2)) for an active suspension system can be derived from general force balance. The nonlinear forces on the tire and the spring are given with Equations (3) and (4) respectively [28], [29]. The active actuator force is given by the Equations (5).

$$m_s \ddot{z}_s + m_s g = -F_{spring} - F_{actuator} + c_o(\dot{z}_s - \dot{z}_u) \tag{1}$$

$$m_u \ddot{z}_u + m_u g = F_{spring} + F_{actuator} - F_{tire} \tag{2}$$

$$F_{tire} = k_{11}(z_u - z_g) + k_{12}(z_u - z_g)^2 - k_{13}(z_u - z_g)^3 \tag{3}$$

$$F_{spring} = k_{21}(z_s - z_u) + k_{22}(z_u - z_g)^3 \tag{4}$$

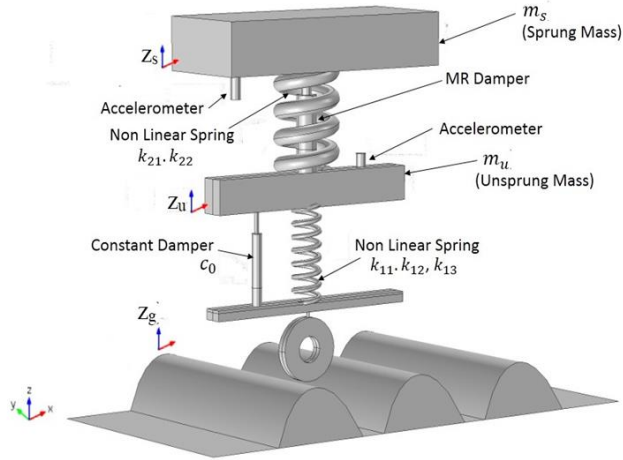

Fig. 1. Semi Active Suspension System

$$F_{actuator} = -b_s \dot{z}_s + b_u \dot{z}_u \tag{5}$$

In (5), $b_s$ and $b_u$ represent the apparent sky hook damper coefficients for the sprung and the unsprung masses, respectively. For a semi active system, the force due to the damper is given by

(6) [30] and it takes into consideration the dissipative constraints of the damper. The equations of motion are given by (7) and (8) respectively.

$$F_{shock} = c_s(\dot{z}_s - \dot{z}_u) + c_i(z_s - z_u) + f_d(t)\tanh\left(k_s(\dot{z}_s - \dot{z}_u) + k_m(z_s - z_u)\right) \tag{6}$$

$$m_s\ddot{z}_s + m_s g = -F_{spring} - F_{shock} + c_o(\dot{z}_s - \dot{z}_u) \tag{7}$$

$$m_u\ddot{z}_u + m_u g = F_{spring} + F_{shock} - F_{tire} - c_o(\dot{z}_s - \dot{z}_u) \tag{8}$$

Hence the equation of the variable damper has the following form:

$$F_{shock} = g_1((\dot{z}_s - \dot{z}_u),(z_s - z_u)) + f_d(t)g_2((\dot{z}_s - \dot{z}_u),(z_s - z_u)) \tag{9}$$

Subject to $0 \leq f_{min} \leq f_d(t) \leq f_{max}$

Where $f_d(t)$ is the control input, which varies accordingly according to the current which generates the magnetic force in the coil of the damping system. The constants $c_i$ and $c_s$ represent the stiffness and the damping coefficients respectively. Here, the variable force $f_d(t)$ has been controlled using a Fuzzy Controller such that it lies within the region bounded by $f_{min}$ and $f_{max}$ and the suspension distortion ($z_s$-$z_u$) has been controlled using a PID controller.

### 3.2 Fuzzy Controller for MR Dampers

A fuzzy controller works on a set of rules that are based on the mathematical analysis and system modelling. The rule-base consists of IF-THEN rule statements depicting conditions for control. The parameters are the input to the "fuzzification" interface where conversion to fuzzy information occurs. After comparison with the fuzzy rule set "defuzzification" converts the conclusions to actual parameters. Figure 2. represents the fuzzy logic control scheme.

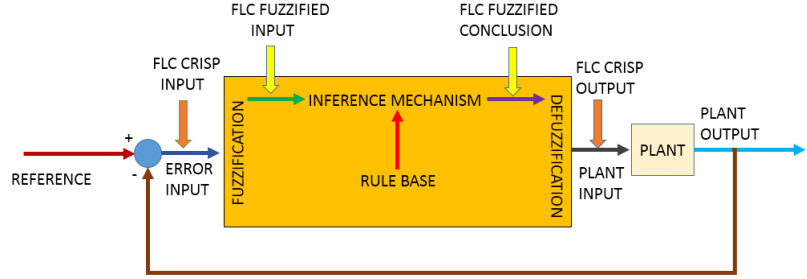

Fig. 2. Fuzzy Controller

Chen [31] developed a skyhook surface sliding mode control method to enhance road comfort in a two degree semi active suspension system. The system utilized a 2-in-1-out FLC rule base having an error variable and a change in error variable as two inputs and the control force (U) as the output variable. A 3-in-1 out FLC system taking sprung mass velocity, the sprung mass acceleration and the suspension distortion as the input has been utilized to calculate the variable term $f_d(t)$ or the output variable, $U,$ for the semi active suspension system in this paper. A generalized bell shaped membership function (MF) (given by Equation10) [32] is a function which defines how the input values are scaled to a membership value between 0 and 1 to facilitate smoothness and differentiability at all points. Asymmetry was introduced into the membership functions by tuning the free parameters (in Equation 10) which has been elucidated in Table 1. Figure 3. shows the definition of the membership for the control output, $U$, while the MFs of the inputs follow the same bell shaped form but with different range of values. In the equation for MF having a variable $x$, the parameters $p$ and $q$ determine the curvature of the MF and $r$ determines the center of the curve .The fuzzy rule set consisted of 9 linguistic notations whose description along with the range of numerical values has been provided below in the Table I.

$$f(x:p,q,r) = \frac{1}{1 + \left|\frac{x-r}{p}\right|^{2q}} \tag{10}$$

TABLE I
FUZZY RULE SET

| Fuzzy Rule Set Notations | Description | Velocity [p;q;r] | Acceleration [p;q;r] | Suspension Distortion [p;q;r] | U [p;q;r] |
|---|---|---|---|---|---|
| NMIN | Negative Minimum | [0.1575, 2.5, -1] | [1.188;2.499 ;-9] | [0.0338;2.5;0.0193] | [21.88 ;2.5; 0] |
| NL | Negative Large | [0.1575, 2.5, 0.685] | [1.188;2.5;0.65] | [0.0376;2.5;0.0944] | [21.88;2.5; 43.75] |
| NM | Negative Middle | [0.1575;2.5;-0.37] | [1.188;2.5;-4.25] | [0.0535;2.2;0.187] | [21.88 ;2.5; 87.5] |
| NS | Negative Small | [0.1575; 2.5; -0.055] | [1.188;2.5;-1.875] | [0.0619;2.5;0.2803] | [21.88 ;2.5; 131.3] |
| Small | Small | [0.1575: 2.5: 0.26] | [1.188;2.5;0.5] | [0.0528;2.5;0.3912] | [21.88 ;2.5; 175] |
| PS | Positive Small | [0.1575;2.5;0.575] | [1.188;2.5;2.875] | [0.0534;2.5;0.4995] | [21.88 ;2.5 ;218.8] |
| PM | Positive Middle | [0.1575;2.5;0.89] | [1.188;2.5;5.25] | [0.0493;2.84;0.6025] | [21.9 ;2.5; 263] |
| PL | Positive Large | [0.1575;2.5;1.205] | [1.188;2.5;7.625] | [0.0403;2.5;0.6972] | [21.88 ;2.5; 306.3] |
| PMAX | Positive Maximum | [0.1575;2.5;1.52] | [1.188;2.501;10] | [0.0458;2.5;0.7815] | [21.88 ;2.5; 350] |

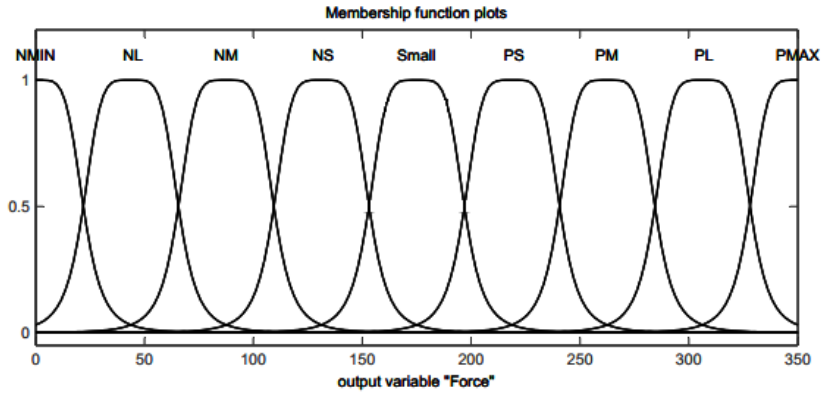

Fig. 3. Membership Function for Control Force Output

The rule base in IF THEN statements based on Mamdani's Fuzzy inference method [33] for (9x9x9) input levels or 729 rule sets was developed considering the competitive nature of the three input variables and the response behavior of the output control force to a change in any of these inputs. It has been observed here that the control force is a strong function of the sprung mass acceleration in the region denoted by $[NM \cup NL \cup NMIN] \cup [PM \cup PL \cup PMAX]$ and is inversely proportional to the acceleration irresepective of the range of values attained by the other two inputs. However in the region denoted by [NS ∪ Small ∪ PS] the control force becomes a strong function of the distortion and the sprung mass velocity and the output is judged on the basis of whose magnitude is greater. The rule base have been given below in a nutshell and the range of the parameter settings along with their definitions has been shown in Table 1.

1. If (Velocity is NL) and (Acceleration is NMIN) and (Distortion is PMAX) then (Force is PL)

2. If (Velocity is NM) and (Acceleration is NMIN) and (Distortion is PMAX) then (Force is PM)

3. If (Velocity is NS) and (Acceleration is NMIN) and (Distortion is NMIN) then (Force is PS)

4. If (Velocity is PS) and (Acceleration is NM) and (Distortion is PL) then (Force is PM)

5. If (Velocity is PM) and (Acceleration is NMIN) and (Distortion is PM) then (Force is PS)

6. If (Velocity is PL) and (Acceleration is NL) and (Distortion is NMIN) then (Force is PL)

7. If (Velocity is NMIN) and (Acceleration is NMIN) and (Distortion is NMIN) then (Force is PMAX)

8. If (Velocity is PL) and (Acceleration is NS) and (Distortion is PL) then (Force is NL)

9. If (Velocity is PL) and (Acceleration is Small) and (Distortion is PS) then (Force is NL)

.

.

726. If (Velocity is NMIN) and (Acceleration is PS) and (Distortion is NM) then (Force is PL)

727. If (Velocity is Small) and (Acceleration is NMIN) and (Distortion is NMIN) then (Force is PS).

728. If (Velocity is PMAX) and (Acceleration is NS) and (Distortion is NMIN) then (Force is NL)

729. If (Velocity is PMAX) and (Acceleration is PMAX) and (Distortion is PMAX) then (Force is NMIN).

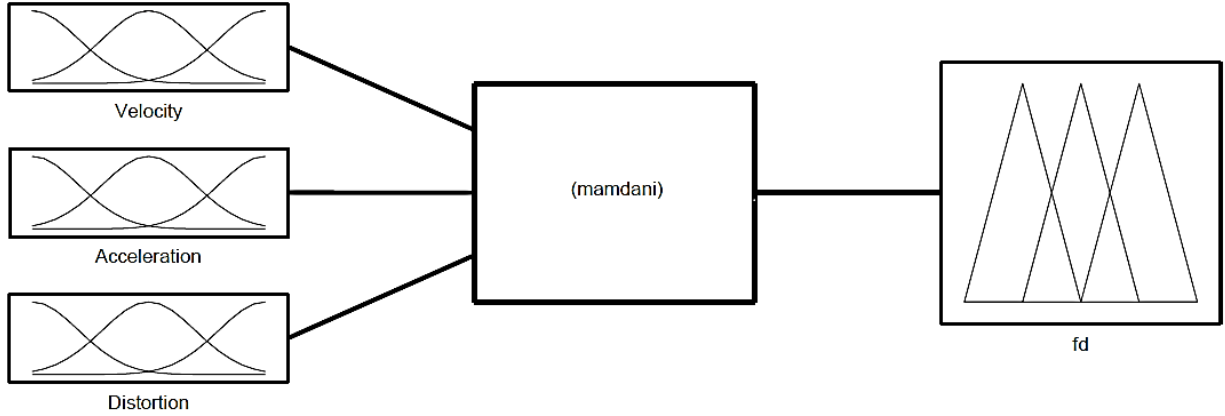

Fig. 4. Fuzzy Logic Control Mechanism

## 3.3. The Fruit Fly Algorithm with PID Tuning

The Fruit Fly algorithm is an optimization procedure, which utilizes the food finding behavior of fruit flies [14]. The FOA generates and initializes a location for the swarm and then each member is assigned a direction. The best position is then estimated on the basis of a judgment function when they reach new positions after an estimation of the distance to the origin and the corresponding Smell which is the reciprocal of the distance. The optimal solution is reached after a series of iterations. There is literature which focusses on the applications of the conventional and modified Fruit Fly algorithms to control PID systems [15, 34]. Here, the error function $e(t)=r(t)-y(t)$ is the input to the PID controller and $u(t)$ is the output of the controller. The actuating signal for PID control is given by,

$$u(t) = k_\mathrm{P} e(t) + k_\mathrm{D} \frac{de(t)}{dt} + k_\mathrm{I} \int e(t) dt \tag{11}$$

The performance of the PID is often characterized by the judgment function that takes into account the aspects of minimizing the error with minimum effort. Given by Equation(12), the piecewise defined function consists of terms which signify the output, error and the overshoot in the PID controller. $w_1, w_2, w_3, w_4$ are the combined weights and $t_p$ is the rising time when the plant output equals the reference input of the PID.

$$\text{minimize} \quad J = \begin{cases} \int_0^\infty \left( w_1 |e(t)| + w_2 u^2(t) \right) dt + w_3 t_p & \text{if } e_{overshoot}(t) > 0 \\ \int_0^\infty \left( w_1 |e(t)| + w_2 u^2(t) + w_4 |e_{overshoot}(t)| \right) dt + w_3 t_p, & \text{otherwise} \end{cases} \quad (12)$$

$$\text{subject to } k_P, k_D, k_I > 0$$

The first function in Equation 12 is valid till the overshoot $e_{overshoot}(t)$ is positive while the second half is valid elsewhere. Based on the minimum judgment value the optimized parameters were calculated.

### 3.4. Chaotic Fruit Fly Algorithm

The Chaotic Fruit Fly Algorithm [11], an improvement on the Fruit Fly Algorithm incorporates chaos [35] into the generation of random data for the purpose of updating locations during the optimization process, in order to solve global optimization problems. The updated location of the parameters is based on logistic chaotic mapping and is given by Equation (13).

$$x(i+1) = x(i) + s x_{balance}(i) + r(1-s) x_{chaos}(i) \quad (13)$$

Where r is a random number and s is a balance parameter that can be chosen from 0 to 1 to enhance possibilities of seeking a global optimum. At s=0 the search is completely independent and at s=1, the search is based on chaotic mapping. After initialization of the parameters like the maximum number of iterations, population size and balance parameter values, chaotic search is initiated. Subsequently, the Distance and Smell Concentration given by Equation (14) are calculated. The Smell Concentration Judgment Value takes into account a trip parameter $\Delta$ which enables the fitness function to be both positive and negative and allows efficient searching of the global minimum for the system.

$$Distance = \sqrt{x^2 + y^2 + z^2} \qquad (14)$$

$$Smell\ Concentration = \frac{1}{Distance} + \Delta;$$

$$\Delta = Distance \times (0.5 - \beta);\ 0 \leq \beta \leq 1$$

For a particular swarm at the end of each iteration the minimum of the Smell Function is calculated and if the smell concentration is higher than the previous value then the new value of concentration is updated after further search. Otherwise the location of the swarm is updated and the new set of values determines the minimum of the function under consideration.

For the purpose of finding the optimal PID gains for the suspension system in this paper, we further extend the Chaotic Fruit Fly algorithm to three dimension where the updated location of the parameters based on logistic chaotic mapping is given by (15) where *x, y and z* basically represent the coordinates of the search space for PID gain parameters; $k_P$, $k_D$, and $k_I$.

$$\left.\begin{array}{l} x(i+1) = x(i) + s x_{balance}(i) + r(1-s) x_{chaos}(i) \\ y(i+1) = y(i) + s y_{balance}(i) + r(1-s) y_{chaos}(i) \\ z(i+1) = z(i) + s z_{balance}(i) + r(1-s) z_{chaos}(i) \end{array}\right\} \qquad (15)$$

### 3.5. PID Tuning Using CFOA

This paper introduces the utility of the Chaotic Fruit Fly algorithm in the tuning of a PID for the control of the suspension distortion. The judgment function defined by (13) and calculated by the closed loop dynamic performance of the suspension system, was minimized after random initiation of the location of PID parameters. Minimization of the judgment function returned the optimized values of the PID parameters. Similar to the CFOA the after initialization of the parameters (STEP 1) and logistics search (STEP 2), the dynamic performance of the PID system has been calculated using the input error and the output variable *u(t)* (STEP 3). After calculation of the Judgment Function in (STEP 4) the minimum of the Judgment Function is calculated (STEP 5) and when the concentration is higher than the value in the previous iteration, the value of PID parameters is updated after further search (STEP 6-STEP 7). Otherwise the location of the swarm is updated and the new set of values determine the minimum of the function under consideration (STEP 8). The operation continues till the maximum number of iterations has reached (STEP 9).The algorithm has been described below in the following Figure 5.

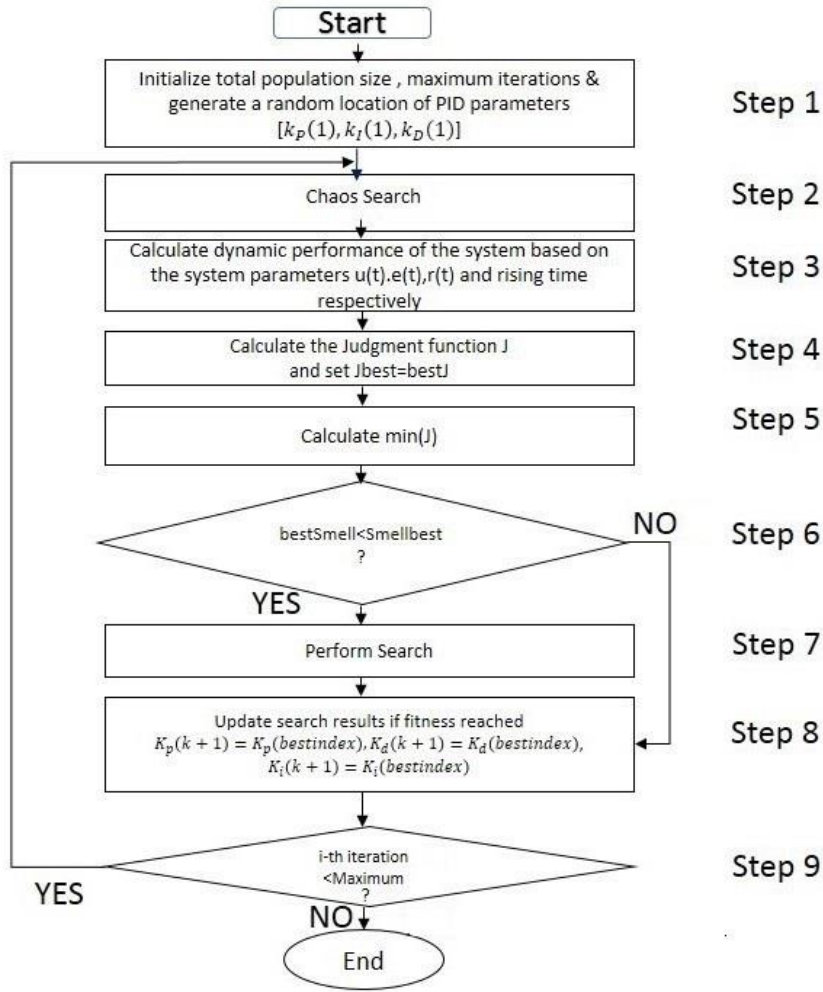

Fig. 5. CFOA in PID Tuning

## 3.6. Road Input

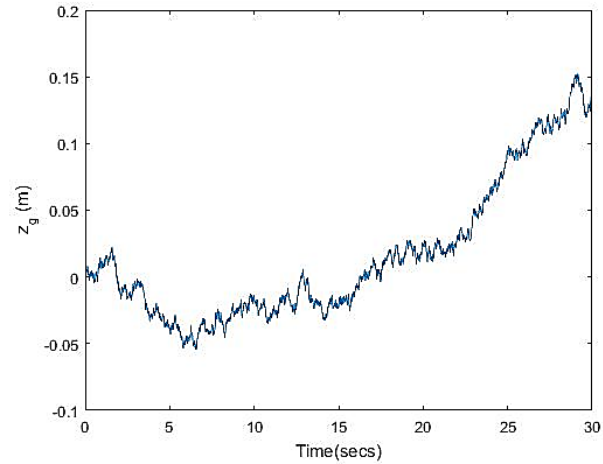

Fig.6. Random Road Profile of roughness class C

A random road profile $z_g$ (Fig.6) was used as an input to the suspension system to test the performance of the CFOA tuned PID Controller. A large number of stochastic road models have been proposed in [36] which introduce variable variations in short sections of the trajectory and can be used for testing the performance of the suspension system considering it to have a unidirectional velocity v in the forward direction. In this study a road profile conforming to ISO standards was considered and was mathematically described as a random function using a Power Spectral Density Function [37]. Based on the road roughness classification proposed by the International Standards Organization the PSD function takes into account the relationship between the Spectral Density and the Spatial Frequency for different road profile classes. A road profile of roughness class C has been considered (16-17) where $C(\Omega_0)$ is the reference spatial frequency characterizing the degree of unevenness and m and n are the constants depicting waviness being equal to 2 and 1.5 respectively.

$$S_{road}(\Omega_i) = C(\Omega_O)(\frac{\Omega_i}{\Omega_0})^{-m}, \quad \Omega_i \leq \Omega_O = \frac{1}{2\pi} cycles/m \qquad (16)$$

$$S_{road}(\Omega_i) = C(\Omega_O)(\frac{\Omega_i}{\Omega_0})^{-n}, \quad \Omega_i > \Omega_O = \frac{1}{2\pi} cycles/\text{n} \qquad (17)$$

The tabulated values of the system parameters have been elucidated below in Table II. The value of the suspension parameters and velocity were taken from [31] and the values of the parameters for variable dampers were taken from [30]. The values of the stiffness and the tire spring constants in Equation 3-4 and the constants of the actuator forces in the active suspension system (Equation 5) was calculated by minimizing the vertical acceleration function as defined in [38]. The block diagram of the semi active suspension system utilizing the CFOA-PID based 3-in-1 out Fuzzy control system has been shown in Figure 7.

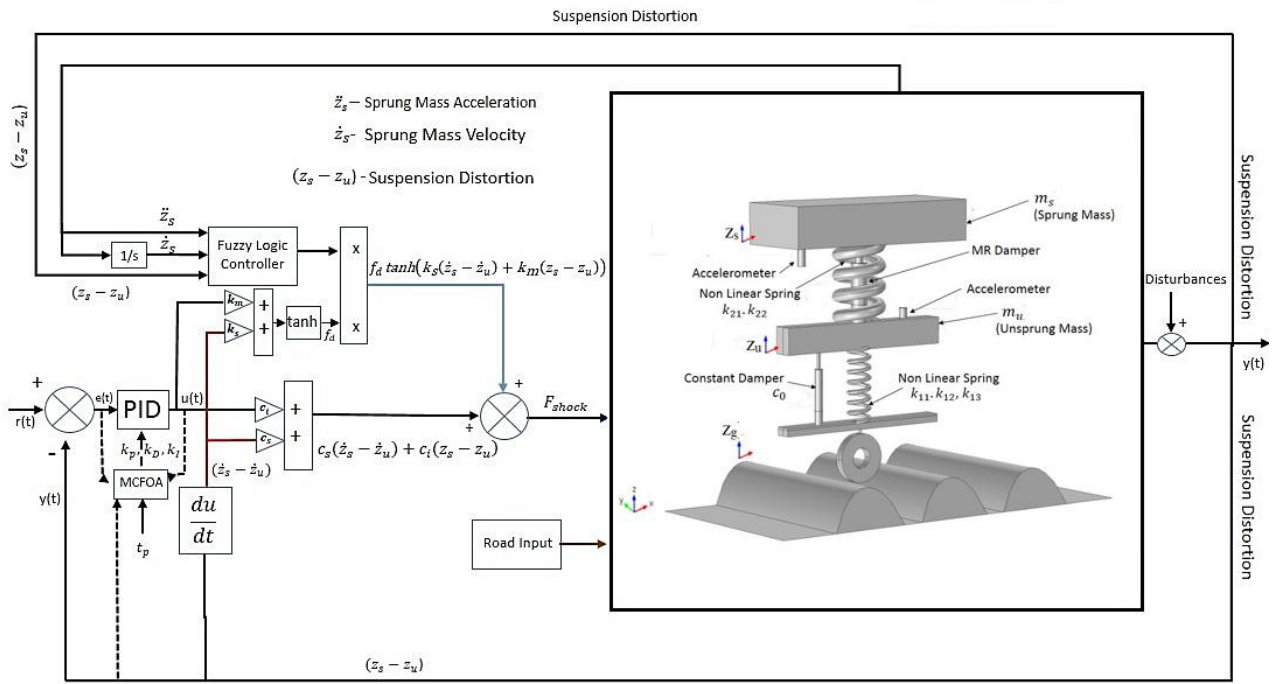

Fig 7. The Block Diagram in SIMULINK

TABLE II
SUSPENSION SYSTEM PARAMETERS

| Parameters | Notation | Value | Unit |
|---|---|---|---|
| Sprung Mass | $m_s$ | 36 | Kg |
| Unsprung Mass | $m_u$ | 240 | Kg |
| Tire Stiffness | $k_{11}$ | 60063 | $Nm^{-3}$ |
|  | $k_{12}$ | 42509 | $Nm^{-2}$ |
|  | $k_{13}$ | 22875 | $Nm^{-2}$ |
| Stiffness of Suspension Damper Coefficients | $k_{21}$ | 15302 | $Nm^{-1}$ |
|  | $k_{22}$ | 2728 | $Nm^{-3}$ |
|  | $c_o$ | 1400 | $N\,m^{-1}s^{-1}$ |
|  | $c_s$ | 620.79 | $N\,m^{-1}s$ |
| Variable Damper Coefficients | $c_i$ | 810.78 | $Nm^{-1}$ |
|  | $k_s$ | 10.54 | $s^{-1}$ |
|  | $k_m$ | 13.76 | $m^{-1}s$ |
|  | $f_{minimum}$ | 0 | N |
|  | $f_{maximum}$ | 350 | N |
| Reference spatial frequency | $\Omega_o$ | 0.1 | $m^{-1}$ |
| Parameters | Notation | Value | Unit |
| Degree of Roughness | $S_{road}(\Omega_i)$ | $256*10^{-6}$ | $m^2/cycles/m$ |
| Unidirectional Velocity | V | 72 | $Km\,h^{-1}$ |
| Active Damper Coefficients | $b_s$ | 1335 | $Nms^{-1}$ |
|  | $b_u$ | 2607 | $Nms^{-1}$ |

## 4. Results and Discussion

The guiding equations of the semi-active suspension system tuned by CFOA based Fuzzy-PID were solved using Simulink$^{TM}$ and based on the optimal values of the PID gain parameters, its performance was judged.The Sprung Mass Acceleration, Tire Loading and the Suspension Distortion for the proposed system were compared with those of the Active System, ZN based PID Fuzzy Hybrid Semi Active System, GA PID Fuzzy Hybrid Semi Active System and similar PID Fuzzy Hybrid Systems tuned by swarm intelligence algorithms like PSO and BFO respectively. Figure 7. shows the fruit fly route to the most optimal solution when the initial conditions of the PID were defined by the conventional ZN tuning. Figure 8. shows the converged minima of the objective function defined by (12).

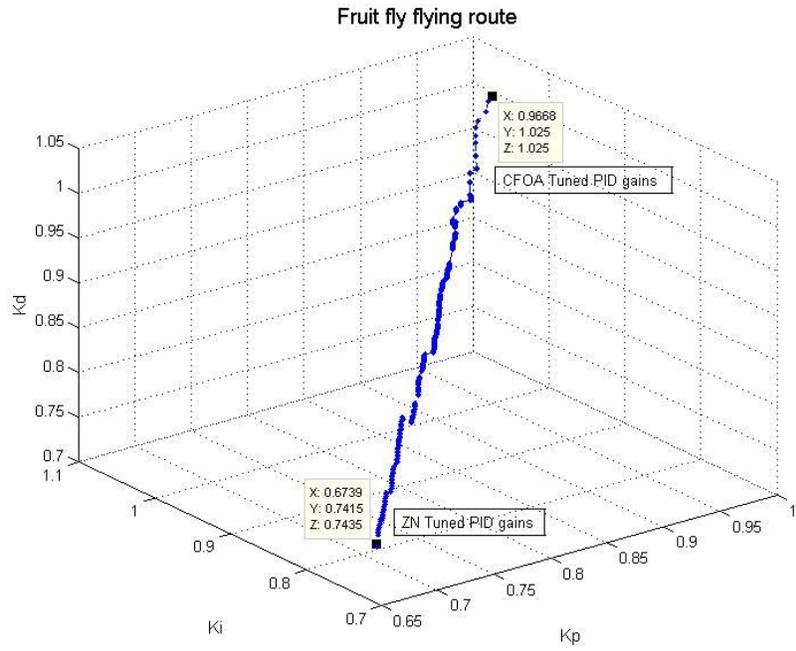

Fig. 8. Fruit Flying Route

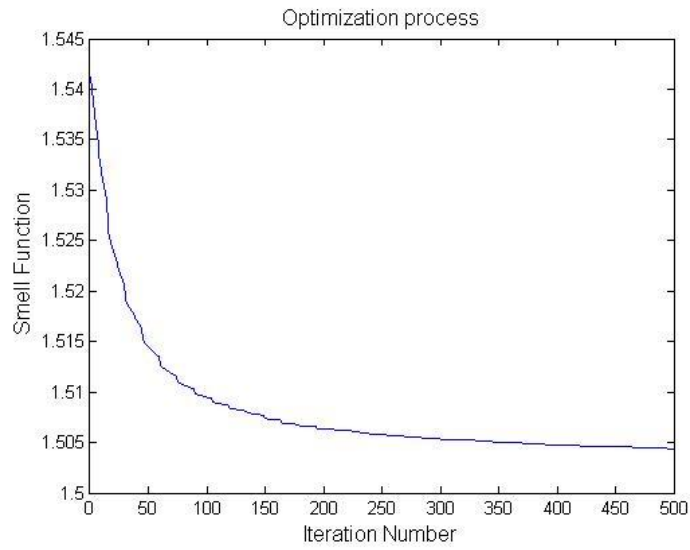

Fig. 9. Minimum value of the Smell Function

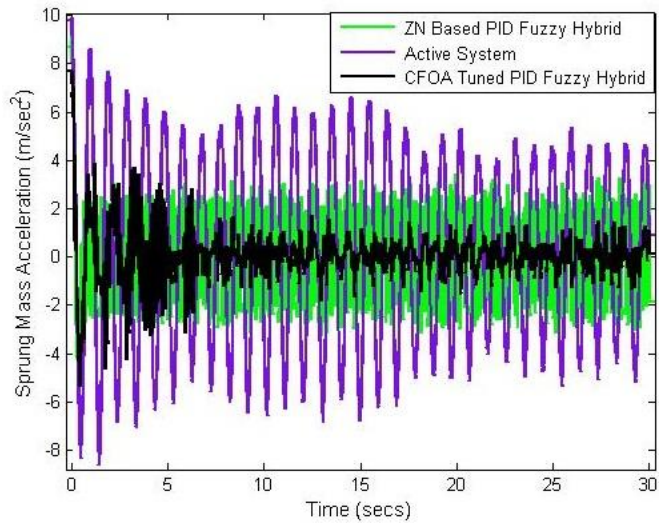

.

Fig. 10. Body Acceleration Profile when compared to
ZN based PID Fuzzy and Active System

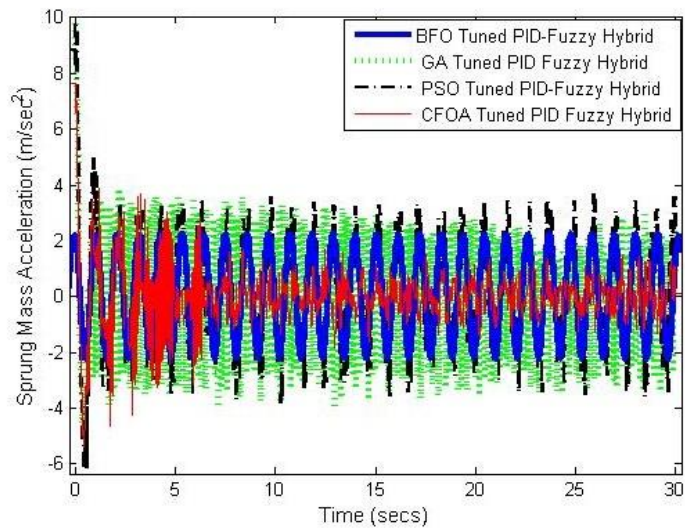

Fig. 11. Body Acceleration Profile when compared to
GA and Swarm Intelligence Tuned PID-Fuzzy Hybrids

Figures 10 and 11 show the variation of the sprung mass acceleration with different types of control strategies in suspension systems. Initially the peaks of the response in the proposed system are high when compared to the other two systems but after a certain period of time they get reduced due to the effective action of the coupled control strategy . Figure 10. shows the vertical body acceleration profile due to changes in the road profile. When compared to the Active System, the maximum peak acceleration got reduced to a maximum of 22% in the case where the CFOA-PID based Fuzzy system was utilized. Additionally the peaks underwent a net reduction of 83.3% when compared to the conventional ZN based PID-Fuzzy Hybrid System. Similarly Figure 11. shows that the CFOA tuned PID Fuzzy Hybrid offers better ride comfort with reduced peaks when compared to the GA and other swarm intelligence algorithms like PSO and BFO respectively. The proposed methodology reduced the maximum peak acceleration by 32.3%, 54.6%, and 25 % when compared to the PSO,BFO and GA based PID Fuzzy Hybrid control strategies, respectively. Thus it can be inferred that the CFOA-PID based Fuzzy System is effective in achieving ride comfort when compared to the Active, ZN-PID based Fuzzy systems, GA and other swarm intelligent systems.

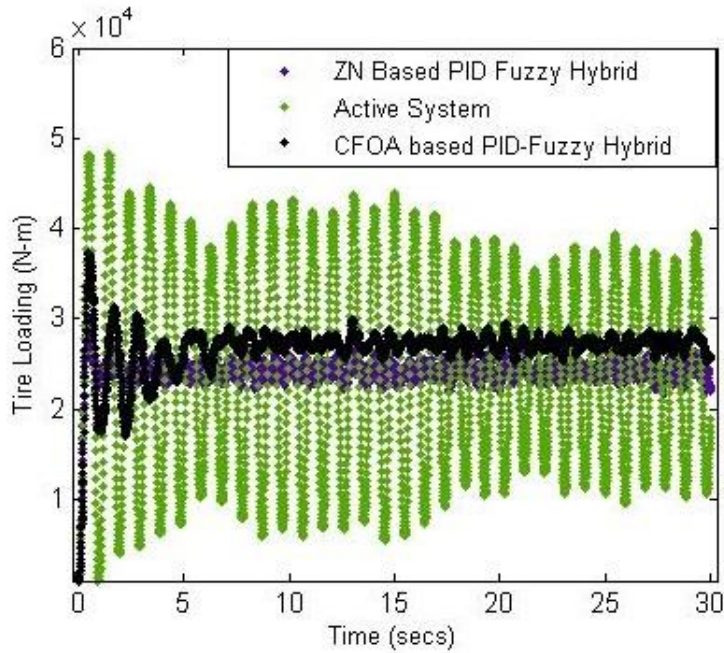

Fig. 12.Tire Loading Profile when compared to
ZN based PID Fuzzy and Active System

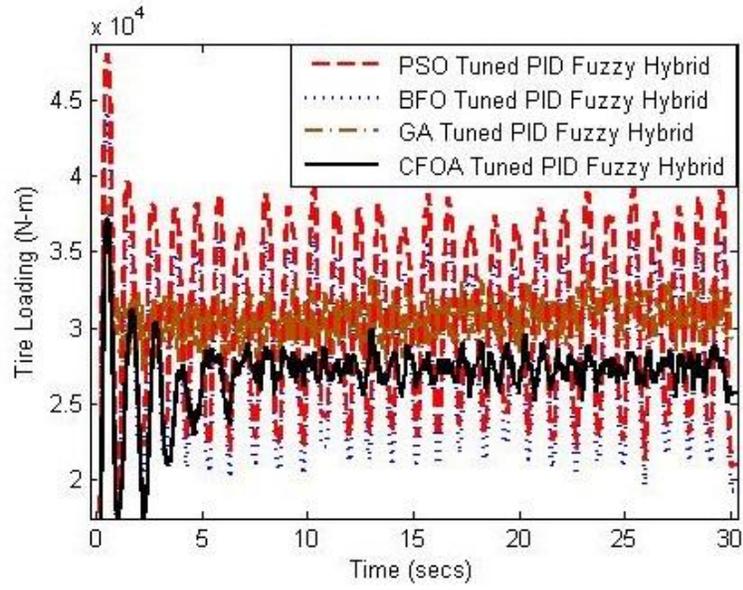

Fig. 13.Tire Loading when compared to
GA and Swarm Intelligence Tuned PID-Fuzzy Hybrids

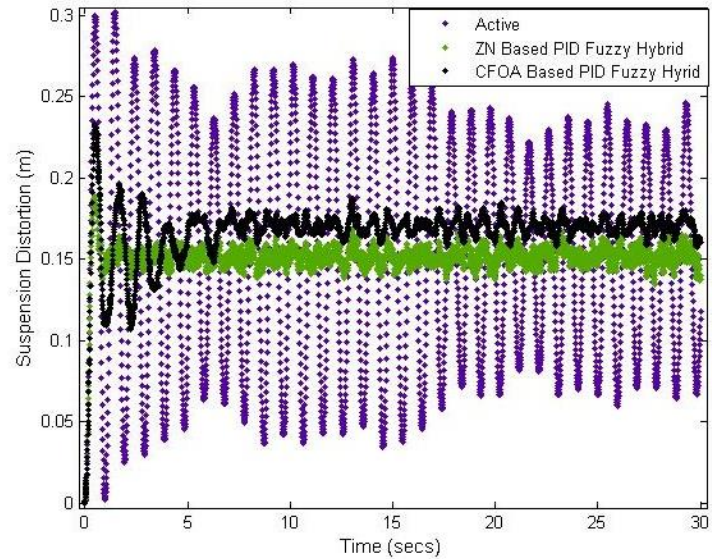

Fig. 14. Suspension Distortion $\left(z_u - z_s\right)$ when compared to
ZN based PID Fuzzy and Active System

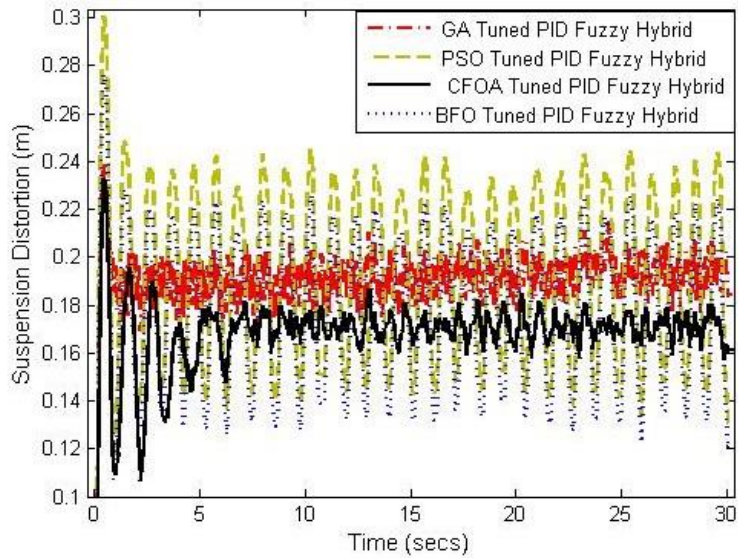

Fig. 15. Suspension Distortion $(z_u - z_s)$ when compared to
GA and Swarm Intelligence Tuned PID-Fuzzy Hybrids

Figures 12, 13, 14 and 15 show the variation of parameters the tire loading and suspension distortion with time which estimate the quality of handling in the vehicular system. In comparison to the Active state the peaks for distortion got reduced to a maximum of 28.57% in the proposed system. The figures also showed that the distortion for the CFOA-PID based Fuzzy system was larger than the ZN based PID Fuzzy system throughout the period of time. This implies that the conventional technique offers better handling when compared to the proposed strategy.

Overall the suspension distortion and tire loading decreases by a maximum of 23.35 %, 22.5% and 5.4 % when compared to the PSO, BFO and GA based systems respectively. Thus compared to the performance of the swarm intelligence algorithms, the proposed system offered better road handling.

5. Conclusiom

A semi active suspension system utilizing a 3-in-1 out Fuzzy PID unit was designed taking into consideration the quarter car model, where the PID was tuned using the Modified Chaotic Fruit Fly Algorithm. The performance of the system was studied using Simulink$^{TM}$ and was compared to the conventional Ziegler Nichols Tuned PID-Fuzzy Hybrid out System, the Active System ,GA and Swarm Intelligence based PID-Fuzzy Systems. The maximum peak values of the body acceleration decreased by 83.3 % when compared to the active system and by 22%, 25%, 32.3% and 54.6% when compared to the Ziegler Nichols, Genetic Algorithms (GA), Particle Swarm Optimization (PSO) and Bacterial Foraging Optimization (BFO) based suspension systems, thereby increasing ride comfort. The suspension distortion and tire loading decreased by 23.35 %, 22.5%, 5.4 % and 28.57% when compared to the PSO, BFO, GA and Active systems respectively. In addition to the controlling action of the PID control system, the linguistic 3-in-1 out rule set of the fuzzy controller took into consideration the effects of distortion, vertical body acceleration and velocity on the variable damping force producing an output which caters to the requirements imposed by the different conditions during transportation. Hence it can be concluded that the Chaotic Fruit Fly Tuned PID-Fuzzy Hybrid strategy based suspension can adapt to the sudden changes in the road profile structure thereby increasing stability during transportation. It offers better performance with respect to swarm intelligent based PID controlled suspension systems when both ride comfort and handling are taken into account.

## 6. Limitation and Scope for Research

The proposed methodology discussed how to increase road comfort and handling in a semi active suspension system utilizing a new PID Fuzzy Hybrid control strategy. Since the PID controls the distortion, its reduction increases handling. Compared to the conventional ZN based PID systems the handling was reduced in the proposed system. There is a need for the development of a control strategy to introduce improved handling when compared to conventional ZN based PID systems. Further research will focus on the introduction of dynamic set point tracking techniques on PID control to further improve road handling compared to conventional suspension systems tuned by ZN based PID systems.

## Nomenclature

| Notation | Parameters |
|---|---|
| $m_s$ | Sprung Mass(Kg) |
| $m_u$ | Unsprung Mass(Kg) |
| $k_{11}$ | Tire Stiffness (Nm$^{-3}$) |
| $k_{12}$ | Tire Stiffness (Nm$^{-2}$) |
| $k_{13}$ | Tire Stiffness (Nm$^{-2}$) |
| $k_{21}$ | Stiffness of Suspension (Nm$^{-1}$) |
| $k_{22}$ | Damper Coefficients(Nm$^{-3}$) |
| $k_m$ | Variable Damper Coefficients (m$^{-1}$s) |
| $f_{minimum}$ | Lower Bound of Damper Force(N) |
| $f_{maximum}$ | Upper Bound of Damper Force(N) |
| $\Omega_o$ | Reference spatial frequency (m$^{-1}$) |
| $S_{road}(\Omega_i)$ | Degree of Roughness (m$^2$/cycles/m) |
| V | Unidirectional Velocity (Km h$^{-1}$) |
| $b_s$ | Active Damper Coefficients(Nms$^{-1}$) |
| $b_u$ | Active Damper Coefficients(Nms$^{-1}$) |
| $c_o$ | Variable Damper Coefficients (N m$^{-1}$s$^{-1}$) |
| $c_s$ | Variable Damper Coefficients (N m$^{-1}$s) |
| $c_i$ | Variable Damper Coefficients (Nm$^{-1}$) |
| $k_s$ | Variable Damper Coefficients (s$^{-1}$) |
| CFOA | Chaotic Fruit Fly Algorithm |
| FLC | Fuzzy Logic Controller |

**Figure Captions List**

Fig.14      Suspension Distortion $(z_u - z_s)$ when compared to GA and Swarm Intelligence Tuned PID-Fuzzy Hybrids

**Table Caption List**

Table I      Fuzzy Rule Set Notations with parameter range settings

Table II      Suspension System Parameters